# Towards Normal Forms for GHZ/W Calculus

Shibdas Roy[1]

*Centre for Quantum Computation and Communication Technology (CQC$^2$T),*
*School of Engineering and Information Technology (SEIT),*
*University of New South Wales (UNSW) at the Australian Defence Force Academy (ADFA),*
*Canberra ACT 2600, Australia.*

**Abstract.** Recently, a novel *GHZ/W graphical calculus* has been established to study and reason more intuitively about interacting quantum systems. The compositional structure of this calculus was shown to be well-equipped to sufficiently express arbitrary mutlipartite quantum states equivalent under stochastic local operations and classical communication (SLOCC). However, it is still not clear how to explicitly identify which graphical properties lead to what states. This can be achieved if we have well-behaved normal forms for arbitrary graphs within this calculus. This article lays down a first attempt at realizing such normal forms for a restricted class of such graphs, namely *simple* and *regular* graphs. These results should pave the way for the most general cases as part of future work.

## 1. INTRODUCTION

A profound interest has developed in studying quantum systems within the mathematical framework of monoidal categories [7] after the phenomenal *Categorical Quantum Mechanics* was first formulated by Abramsky and Coecke in the Oxford University Computing Laboratory [1]. It was further shown by Coecke in [3] that the graphical language(s) of monoidal categories [10] can be suitably exploited to study quantum phenomena more intuitively as compared to the counter-intuitive Hilbert-space formalism. Pragmatically, it was demonstrated that the otherwise very complicated and involved algebraic calculations boil down to much simpler and intuitive manipulations in the graphical paradigm, which Coecke refers to as *Kindergarten Quantum Mechanics* [2]. These works mainly dealt with representing quantum protocols within the categorical regime and reason about them in the related graphical setting.

However, obtaining a generic structural understanding of arbitrary *N*-qubit quantum states is a crucial open problem in quantum computer science, given the central role played by multipartite entanglement in various quantum protocols. It has recently been elucidated in [5] that multipartite quantum entanglement admits a very well-behaved compositional structure within the abstract setting of commutative Frobenius algebras expressed internal to symmetric monoidal categories. In particular, a powerful graphical GHZ/W calculus has been established that allows for axiomatic underpinning in composing arbitrary multipartite quantum states within the symmetric monoidal category **FdHilb**, that has finite-dimensional Hilbert spaces as objects, linear maps as morphisms and the tensor product as the monoidal structure. This calculus was also shown to have refined the graphical calculus of complementary observables [4], which was already previously shown to have many applications and admit automation. The GHZ/W calculus has now been shown in [6] to allow for faithfully encoding standard rational arithmetic as well.

The purpose of this article is to make a first attempt at solving the non-trivial problem of laying down normal forms for arbitrary graphs within the GHZ/W calculus, in particular spelling them out for a restricted class of such graphs, namely *simple* and *regular* graphs. The exhaustive toolkit to express any arbitrary graph in a normal form would naturally allow us to explicitly identify which graphical properties lead to what states and how such states may be classified for arbitrary number of qubits.

In section (2), we set the required background, mainly introducing commutative Frobenius algebras (CFAs) internal to symmetric monoidal categories (SMCs). We discuss the variants of CFAs and their correspondence to the two kinds of tripartite states, viz. GHZ and W states. In section (3), we briefly discuss the GHZ/W calculus, as first laid down

---

in [5]. The core results pertaining to normal forms for *simple* and *regular* graphs are then outlined in section (4). We conclude by summarising our results and making remarks about directions for future work in section (5).

## 2. BACKGROUND

**Definition 2.1.** A Commutative Frobenius Algebra (CFA) internal to a symmetric monoidal category **C** is an object $F$ and four maps $\mu : F \otimes F \to F$, $\eta : \mathbf{I} \to F$, $\delta : F \to F \otimes F$ and $\varepsilon : F \to \mathbf{I}$, such that

- $(F, \mu, \eta)$ is an internal commutative monoid,
- $(F, \delta, \varepsilon)$ is an internal cocommutative comonoid,
- $(\mu \otimes 1_F) \circ (\delta \otimes 1_F) = \delta \circ \mu$

The graphical counterparts for $\mu, \eta, \delta, \varepsilon$ are $\curlyvee, \uparrow, \curlywedge, \downarrow$, respectively, such that we can define:

$$cap := \quad = \quad = \quad \text{and} \quad cup := \quad = \quad = \quad$$

Graphically, the third condition above (*Frobenius law*) is:

$$\quad = \quad$$

Any connected CFA-morphism admits the normal form(s), respecting the number of inputs, outputs and loops:

$$\quad = \quad \tag{1}$$

A *special* CFA (SCFA) and an *anti-special* CFA (ACFA) are respectively defined such that:

$$\quad = \quad \qquad \quad = \quad$$

where $\bigcirc = \cup \circ \cap$. In **FdHilb**, $\bigcirc = D$, the dimension of the underlying Hilbert space, and $\ominus = \frac{1}{D}$.

**Theorem 2.2.** *[5]* **SCFAs are GHZ states and ACFAs are W states:** *Each SCFA $\mathcal{G}$ (respectively ACFA $\mathcal{W}$) on $\mathbb{C}^2$ in **FdHilb** canonically induces a symmetric state in $\mathbb{C}^2 \otimes \mathbb{C}^2 \otimes \mathbb{C}^2$ which is SLOCC-equivalent to $|GHZ\rangle$ (respectively $|W\rangle$). Conversely, any symmetric state that is SLOCC-equivalent to $|GHZ\rangle$ (respectively $|W\rangle$) arises from a unique SCFA $\mathcal{G}$ (respectively ACFA $\mathcal{W}$) on $\mathbb{C}^2$ in **FdHilb**.*

## 3. GHZ/W CALCULUS

The induced CFAs in **FdHilb** for GHZ and W states are:

**GHZ-structure:**

$$\curlyvee = |0\rangle\langle 00| + |1\rangle\langle 11| \qquad \uparrow = \sqrt{2}|+\rangle := (|0\rangle + |1\rangle)$$
$$\curlywedge = |00\rangle\langle 0| + |11\rangle\langle 1| \qquad \downarrow = \sqrt{2}\langle +| := (\langle 0| + \langle 1|) \tag{2}$$

**W-structure:**

$$\curlyvee = |1\rangle\langle 11| + |0\rangle\langle 01| + |0\rangle\langle 10| \qquad \uparrow = |1\rangle$$
$$\curlywedge = |00\rangle\langle 0| + |01\rangle\langle 1| + |10\rangle\langle 1| \qquad \downarrow = \langle 0| \tag{3}$$

Note that the cups and caps induced by each CFA do not coincide:

$$|10\rangle + |01\rangle = \;\triangle\; \neq \;\triangledown\; = |00\rangle + |11\rangle$$

It can be easily verified that:

$$\text{(GHZ tree)} = \text{(GHZ tree)} = |000\rangle + |111\rangle = |GHZ\rangle \qquad \text{(W tree)} = \text{(W tree)} = |100\rangle + |010\rangle + |001\rangle = |W\rangle$$

**Definition 3.1.** [5] A *GHZ/W-pair* consists of a SCFA $(\curlyvee, \circ, \curlywedge, \circ)$ and an ACFA $(\blacktriangledown, \bullet, \blacktriangle, \bullet)$ which satisfy the following four equations.

(i.) $\;\big|\;:= \;\cup\!\curlyvee\; = \;\cup\!\blacktriangledown\;$ \qquad (ii.) $\;\blacktriangle\; = \;\curlywedge\;$

(iii.) $\;\blacktriangle\; = \;\bullet\;\bullet\;$ \qquad (iv.) $\;\bigcirc\!\bullet\; = \;\circ\;$

It can be easily seen that the 'tick' is just the NOT-gate, i.e. a classical structure:

$$\big| = \begin{pmatrix} 0 & 1 \\ 1 & 0 \end{pmatrix}$$

## 4. NORMAL FORMS

We refer to a general graph composed of a GHZ/W pair as a *GHZ/W-graph*. Also, we shall henceforth refer to a graph containing only GHZ nodes (respectively W nodes) and ticks as a *GHZ-graph* (respectively *W-graph*).

**Definition 4.1.** We define an arbitrary connected GHZ/W-graph as **simple**, if and only if it is possible to traverse from every node to every other node, in the graph through edges without a tick.

In other words, if upon removing all ticked edges from a graph, it is still connected, it is referred to as simple.

**Definition 4.2.** We define an arbitrary connected GHZ/W-graph as **regular**, if and only if it is possible to traverse from every GHZ node (respectively W node) to every other GHZ node (respectively W node) in the graph through edges (with or without a tick) without touching a W node (respectively GHZ node).

In other words, if upon removing all the edges connecting unlike nodes, i.e. a GHZ node and a W node, from a graph, it gets split into two graphs, one with all GHZ nodes and the other with all W nodes, it is referred to as regular.

**Remark 4.3.** Note that the tick itself is composed of both GHZ and W nodes, but it essentially yields a classical structure as mentioned before, so we do not count the inherent GHZ or W nodes of a tick above.

The following graphical lemmas hold in **FdHilb**.

**Lemma 4.4.** *[9] Bialgebra rules:*

$$\text{(diagram)} \;\underline{\overset{\beta_1}{=}}\; \text{(diagram)} \qquad \text{(diagram)} \;\underline{\overset{\beta_2}{=}}\; \text{(diagram)} \qquad \text{(diagram)} \;\underline{\overset{\beta_3}{=}}\; \text{(diagram)}$$

**Lemma 4.5.** $\;\phi\; := \;\big|\; := zero\;map$

*Proof.* Plugging $\circ$ and $\bullet$ into the input, we get respectively (Recall $\;\bullet\; = 0$):

$$\text{(diagram)} = \text{(diagram)} = \text{(diagram)} \qquad \text{(diagram)} = \text{(diagram)} = \text{(diagram)}$$

□

## 4.1. Simple GHZ- or W-graphs

**Theorem 4.6.** *Any simple connected GHZ-graph or W-graph admits the normal form given by:*

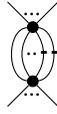

*Proof.* Let us denote the given *simple* GHZ-graph or W-graph as $H$.

If $H$ has no tick, then it admits the normal form of Eq. (1).

If it has one tick, the tick can be removed from the rest of the graph, which is then without any ticks at all.

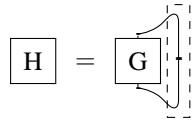

The remaining graph $G$ has, therefore, an additional input and an additional output but no ticks anymore and, thus, admits the normal form of Eq. (1).

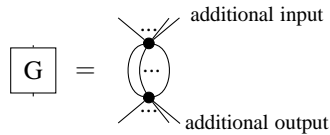

Now, including the tick back into the normalised graph $G$, we get the graph that is normalised and equivalent to the original graph $H$.

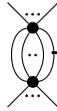

By induction, we can always rewrite a *simple* GHZ-graph or W-graph to the normal form:

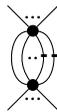

□

*Simple GHZ-graphs.* Every simple connected GHZ-graph is uniquely determined by its number of inputs, number of outputs, number of loops and/or the number of ticks, such that it is either equal to a spider or a **zero map** in **FdHilb**.

*Simple W-graphs.* Every simple connected W-graph is uniquely determined by its number of inputs, number of outputs and the number of loops less the number of ticks.

## 4.2. Regular GHZ/W-graphs

**Lemma 4.7.** 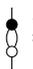 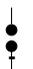 and 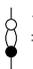

*Proof.* We shall only prove the first here, and the second naturally follows by *dualising* the first.

Applying Eq. ($\beta_2$), we get

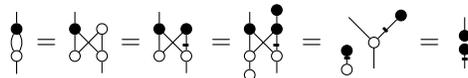

□

**Lemma 4.8.** 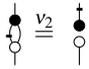 and 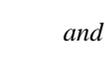

*Proof.* We shall only prove the first here, and the second naturally follows by *dualising* the first.

Applying Eq. ($\beta_3$), we get

□

**Lemma 4.9.**

*Proof.* The proof is simple as shown below:

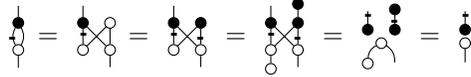

□

**Theorem 4.10.** *Any regular GHZ/W-graph admits the normal form given by:*

*where* **W** *is a W-graph,* **G** *is a GHZ-graph, whereas* **M₁** *and* **M₂** *are respectively of the (mixed) forms:*

*and* 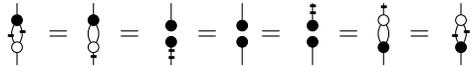

*Proof.* The proof is similar to that for theorem (4.6). In this case, we first inductively remove all the edges (with or without ticks) connecting unlike nodes, i.e. GHZ-node and W-node, resulting in two disconnected parts, one being a GHZ-graph and the other being a W-graph. Note that each such edge removed gives rise to an additional input for one of the two parts and an additional output for the other.

If either or both of the GHZ- and W-graphs are *simple*, we can express them in the normal form of theorem (4.6). We can then add all the edges back, such that the resulting graph is in the above normal form. The edges that were dropped in the first place, upon inclusion into the graph, give rise to the mixed morphisms **M₁** and **M₂**. □

It remains now to explore the nature/behavior of the graphs **M₁** and **M₂** above.

**Theorem 4.11.** *(See proof in Appendix A) A mixed morphism of the form* **M₁***:*

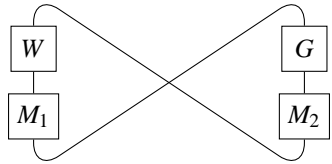

*reduces to simple disconnected graphs or zero morphisms in* **FdHilb** *according to the following rules (where t is the number of ticks and l is the number of loops):*

1. *When $t = 0$, then the morphism always equals the following irrespective of the number of loops:*

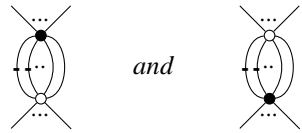

2. When $t = 1, t < l$, then the morphism always equals the following irrespective of the number of loops:

3. When $t = l + 1$, then the morphism always equals the following irrespective of the number of loops:

4. When $t = l, l > 1$, then the morphism always equals the following irrespective of the number of loops:

5. When $t = l = 1$, then the morphism is just the following:

6. When $1 < t < l$, then the morphism always equals zero morphism irrespective of the number of loops.

**Corollary 4.12.** *The reduced forms for various cases for a mixed morphism of the form of* $\mathbf{M_2}$ *are just upside-down of those in Theorem (4.11), since the morphism is obtained by just dualising the one treated in the above theorem.*

## 5. CONCLUSION AND FUTURE WORK

We studied and laid down the normal forms admitted by certain subsets of arbitrary graphs within the GHZ/W-calculus. Such normal forms allow for suitably classifying arbitrary multipartite quantum states and provide an insight into the kind of graphical properties that lead to given states. Precisely, two otherwise inequivalent graphs may be equivalent upon expressing them in the normal forms, such that they essentially represent the same class of quantum states.

However, our treatment in this work has been limited to only so-called 'simple' and 'regular' graphs. Nevertheless, one can always reduce parts of any general graph into regular ones and parts of each GHZ- and W-graphs in each regular piece into simple ones and, in turn, reason about the overall graph upon expressing them in their respective normal forms. We hope to obtain an exhaustive normal form as part of future work for the most general case to be able to completely qualify all known multipartite states more effectively than piecewise reasoning in this language.

## ACKNOWLEDGMENTS

The author is grateful to his MSc thesis supervisor Bob Coecke for his expert guidance and supervision.

## A. PROOF FOR THEOREM 4.11

*Proof.* We prove the different cases one by one by induction below.

1. For $l = 1$, it follows directly from lemma ($v_1$).
   For $l = L > 1, L = even$, applying lemma ($v_1$) to $\frac{L}{2}$ loops the morphism reduces to

   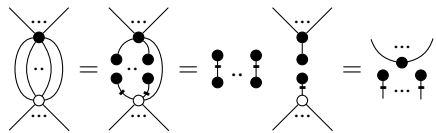

   The number of copies of ⦁ above is equal to $\frac{L}{2} - 1$ and they disappear because ⦁ $= 1_I$. The last step uses the fact that a black dot is copied by a white dot.
   Now, for $l = L + 1$, $l$ is odd, such that the morphism reduces as follows

   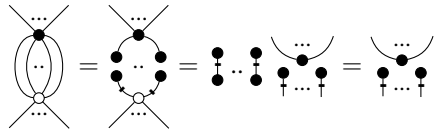

   The number of copies of ⦁ above is equal to $\frac{L}{2}$.
   It can be easily verified that when $L = odd$ the reduction is just vice-versa for $l = L$ and $l = L + 1$ as compared to the case $L = even$.

2. For $t = 1, t < l$, the starting point for induction is $l = 2$.
   For $l = 2$, applying lemma ($v_2$) we get

   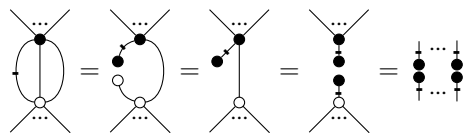

   The last step uses the fact that a black dot is copied by a white dot and that ⦁ $=$ ⦁ ⦁.
   For $l = L > 2, L = even$, applying lemma ($v_2$) to the leftmost loop and lemma ($v_1$) to the next $\frac{L}{2} - 1$ loops, the morphism reduces to

   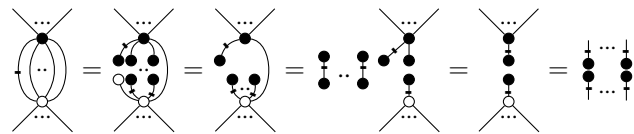

   The number of copies of ⦁ above is equal to $\frac{L}{2} - 1$.
   Now, for $l = L + 1$, $l$ is odd, such that the morphism reduces as follows

   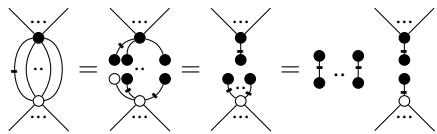

The number of copies of ![symbol] above is equal to $\frac{L}{2}$.

It can be easily verified that when $L = odd$ the reduction is just vice-versa for $l = L$ and $l = L+1$ as compared to the case $L = even$.

3. For $l = 1$, it follows directly from lemma $(v_3)$.

    For $l = L > 1, L = even$ applying lemma $(v_3)$ to $\frac{L}{2}$ loops the morphism reduces to

    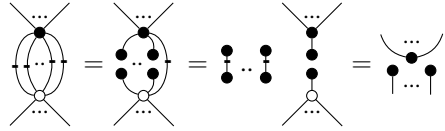

    The number of copies of ![symbol] above is equal to $\frac{L}{2} - 1$.

    Now, for $l = L+1$, $l$ is odd, such that the morphism reduces as follows

    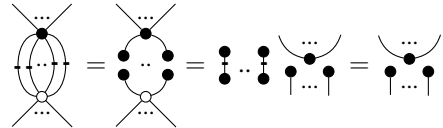

    The number of copies of ![symbol] above is equal to $\frac{L}{2}$.

    It can be easily verified that when $L = odd$ the reduction is just vice-versa for $l = L$ and $l = L+1$ as compared to the case $L = even$.

4. For $t = l, l > 1$, the starting point for induction in this case is $l = 2$.

    For $l = 2$, applying lemma $(v_3)$ to the left loop we get

    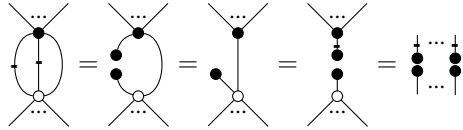

    For $l = L > 2, L = even$, applying lemma $(v_3)$ to the leftmost $\frac{L}{2}$ loops, the morphism reduces to

    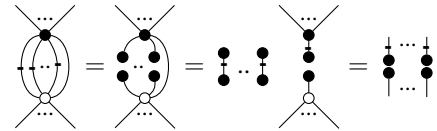

    The number of copies of ![symbol] above is equal to $\frac{L}{2} - 1$.

    Now, for $l = L+1$, $l$ is odd, applying lemma $(v_3)$ to the left $\frac{L}{2}$ loops and lemma $(v_2)$ to the rightmost loop, the morphism reduces as follows

    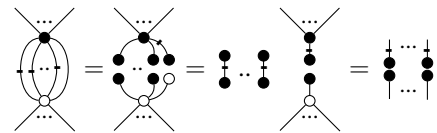

    The number of copies of ![symbol] above is equal to $\frac{L}{2}$.

    It can be easily verified that when $L = odd$ the reduction is just vice-versa for $l = L$ and $l = L+1$ as compared to the case $L = even$.

5. This follows directly from lemma $(v_2)$.

6. It can be easily verified that in this case, the morphism can always be reduced to an arbitrary disconnected graph multiplied by copies of ![symbol], which is zero in **FdHilb**, thereby reducing the morphism to a zero morphism.

□